\documentclass[a4paper]{jpconf}
\usepackage{graphicx}
\usepackage{cite} 
\def\lsim{\raise0.3ex\hbox{$\;<$\kern-0.75em\raise-1.1ex\hbox{$\sim\;$}}}
\def\gsim{\raise0.3ex\hbox{$\;>$\kern-0.75em\raise-1.1ex\hbox{$\sim\;$}}}
\newcommand{\bino}{\widetilde{\cal B}}
\begin{document}
\title{How light can the lightest neutralino be?}

\author{Olaf Kittel}

\address{Departamento de F\'isica Te\'orica y del Cosmos and CAFPE, \\
Universidad de Granada, E-18071 Granada, Spain
}

\ead{kittel@th.physik.uni-bonn.de}

\begin{abstract}
In this talk we summarize previous work on mass bounds
of a light neutralino in the  Minimal Supersymmetric Standard Model.
We show that without the GUT relation between the gaugino mass parameters 
$M_1$ and $M_2$, the mass of the lightest neutralino is essentially 
unconstrained by collider bounds  and precision observables.
We conclude by considering also the astrophysics and cosmology of 
a  light neutralino.
\end{abstract}

\section{Introduction}
%lala
\medskip
The lightest supersymmetric particle, the LSP, plays a special role in
the search for Supersymmetry (SUSY)~\cite{Haber:1984rc} at colliders. 
For conserved R-parity or proton hexality~\cite{Dreiner:2005rd,Dreiner:2007vp},
the LSP is stable and thus the end product of cascade decays of any produced 
SUSY particle. Thus the nature of the LSP is decisive for all supersymmetric 
signatures at the LHC and ILC.  Here we ask the question
`How light can the lightest neutralino be?', and discuss bounds from collider 
physics and precision observables, to summarize previous 
works~\cite{Choudhury:1999tn,Dreiner:2003wh,Dreiner:2006sb,Dreiner:2007vm,Dreiner:2009ic,Dreiner:2009yk}.
Note that over the last decade there has been tremendous
interest to derive bounds on the neutralino mass mainly from its relic density 
to explain the dark matter of the universe~\cite{cosmology,Bottino:2009km,Vasquez:2010ru,Feldman:2010ke}. 

\section{Neutralino framework}
\medskip

In the Minimal Supersymmetric Standard Model (MSSM)~\cite{Haber:1984rc},
the masses and mixings of the neutralinos and charginos are given by their 
mass matrices~\cite{Haber:1984rc,Amsler:2008zz}
\begin{equation}
%\tiny
{\mathcal M}_0
= M_Z\!\left(\!\!\begin {array}{cccc} M_1/M_Z&0&
          -s_\theta c_\beta &
\phantom{-}s_\theta s_\beta  \\
0&M_2/M_Z& 
\phantom{-}c_\theta c_\beta  &
          -c_\theta s_\beta \\
          -s_\theta c_\beta  &
\phantom{-}c_\theta c_\beta  &0&-\mu/M_Z\\
\phantom{-}s_\theta s_\beta &
-c_\theta s_\beta &-\mu/M_Z&0
\end{array}
\!\!\right), \quad
{\mathcal M}_\pm=M_W\!\left(\!\!
\begin{array}{cc}
M_2/M_W &\sqrt{2} s_\beta\\
\sqrt{2}c_\beta&\mu/M_W 
\end{array}
\!\!\right), 
\end{equation}
respectively, with $c_\beta= \cos\beta$, $s_\beta= \sin\beta$ and
$c_\theta=\cos\theta_w$, $s_\theta=\sin\theta_w$,
and the weak mixing angle $\theta_w$.
Besides the masses  of the $W$ and $Z$ boson, $M_W$ and $M_Z$, respectively,
the neutralino and chargino sectors at tree level only depend on the 
$U(1)_Y$ and $SU(2)_L$ gaugino masses $M_1$ and $M_2$, respectively,
the higgsino mass parameter $\mu$, and the ratio $\tan\beta=v_2/v_1$ of the 
vacuum expectation values of the two Higgs fields.
The neutralino (chargino) masses are the square roots of the eigenvalues of
${\mathcal M}_0 { \mathcal M}_0^\dagger$ 
(${\mathcal M}_\pm {\mathcal M}_\pm^\dagger$)~\cite{Amsler:2008zz}.

%\medskip
\newpage

The PDG cites as the laboratory bound on the lightest 
neutralino mass~\cite{Amsler:2008zz}
\begin{equation}
m_{\tilde\chi_1^0}>46~{\rm GeV}    %,\quad @\, 95\%\;\mathrm{C.L.}\,,
\label{lep-bound}
\end{equation}
at $95$\% C.L., which is based on the chargino searches at LEP,
$m_{\tilde\chi^\pm_1}\gsim100$~GeV~\cite{Amsler:2008zz}. These
yield lower limits on  $M_2,|\mu| \gsim 100$~GeV.  Furthermore, this bound
assumes an underlying SUSY GUT, i.e,  
 $M_1 = 5/3 \tan^2(\theta_w)M_2\approx 0.5\, M_2$.
The experimental bound on $M_2$  then implies  the lower bound on $M_1$,
which give rise to the lower bound in Eq.~(\ref{lep-bound}).

%%%%%%%%%%%%%%%%%%%%%%%%%%%%%%Fig 1%%%%%%%%%%%%%%%%%%%%%%%%%%%%%%%%%%%%%%%%%%%%%%%%
\begin{figure}[t]
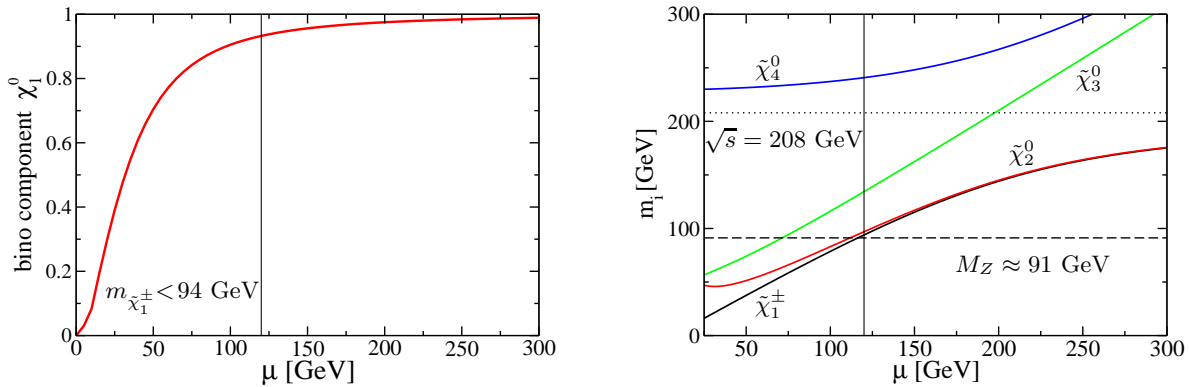

\begin{picture}(200,80)
\put(-5,-10){\includegraphics{./N11mu_200.eps}}
\put(230,-10){\includegraphics{./plot_masses.eps}}
\put(32,30){\footnotesize $m_{\tilde\chi_1^\pm}\!\!<\!94$~GeV}
\put(370,82){\footnotesize $\tilde\chi_2^0$}
\put(395,110){\footnotesize $\tilde\chi_3^0$}
\put(275,114){\footnotesize $\tilde\chi_4^0$}
\put(275,25){\footnotesize $\tilde\chi_1^\pm$}
\put(256,87){\footnotesize $\sqrt s= 208$~GeV}
\put(350,40){\footnotesize $M_Z\approx 91$~GeV}
\end{picture}
\caption{\small 
        Bino admixture of $\tilde\chi_1^0$ (left plot) and masses of 
        charginos and neutralinos (right plot)
        for $M_2=200$~GeV, $\tan\beta=10$,
        and $M_1$ as given in Eq.~(\ref{Eq:zeromass}), such that
        $m_{\tilde\chi_1^0}=0$~GeV~\cite{Dreiner:2009ic}.
        Left to the vertical lines at $\mu\approx135$~GeV, the chargino
        mass is $m_{\tilde\chi_1^\pm}< 94$~GeV.
        In the right panel, the dotted line indicates the reach of LEP2
        ($\sqrt s= 208$~GeV)
        for $e^+e^- \to \tilde\chi_1^0\tilde\chi_i^0$ production,
        and the dashed line indicates
        the mass of the Z boson,  $M_Z\approx 91$~GeV.
}
\label{fig:neutmixandmass}
\end{figure}
%%%%%%%%%%%%%%%%%%%%% %%%%%%%%%%%%%%%%%%%%%%%%%%%%%%%%%%%%%%%%%%%%%%%%%%%%%%

\medskip

However, if one drops the GUT relation, $M_1$ is an independent
parameter, allowing to tune the neutralino mass 
determined from the lowest-order mass matrix ${\mathcal M}_0$
freely~\cite{Choudhury:1999tn,Gogoladze:2002xp,cosmology,Dreiner:2009ic}.
This choice can be made stable against radiative corrections~\cite{Dreiner:2009ic}.
 The neutralino  mass is identically zero for~\cite{Gogoladze:2002xp}
\begin{eqnarray}
{\rm det}({\mathcal M}_0)=0\quad \Rightarrow M_1 = 
\frac{M^2_Z M_2 \sin^2\theta_w \sin(2\beta)}
{\mu M_2 - M_Z^2  \cos^2\theta_w \sin(2\beta)}
%\approx \frac{m_Z^2}{\mu} \sw[2]\sin(2\beta) 
\approx 0.05 \frac{M_Z^2}{\mu} = \mathcal{O}(1\,\mathrm{GeV}). 
\label{Eq:zeromass}
\end{eqnarray}
For $M_1\ll M_2,|\mu|$, the neutralino $\tilde\chi^0_1$ is mainly a bino,
see  Fig.~\ref{fig:neutmixandmass}, i.e., it couples to hypercharge.
This will automatically reduce the contribution to the invisible $Z$ width, 
$Z\to \tilde\chi^0_1\tilde\chi^0_1$.
The masses of the other neutralinos and charginos are then of the order of 
$M_2$ and $|\mu|$, see Fig.~\ref{fig:neutmixandmass}. 
In the following, we discuss bounds on the neutralino mass
from production at LEP and from precision observables.
Finally, we summarize bounds from astrophysics and cosmology.

\section{Collider bounds}

\medskip

\begin{figure}[t]
\begin{picture}(200,150)
\put(0,-20){\includegraphics{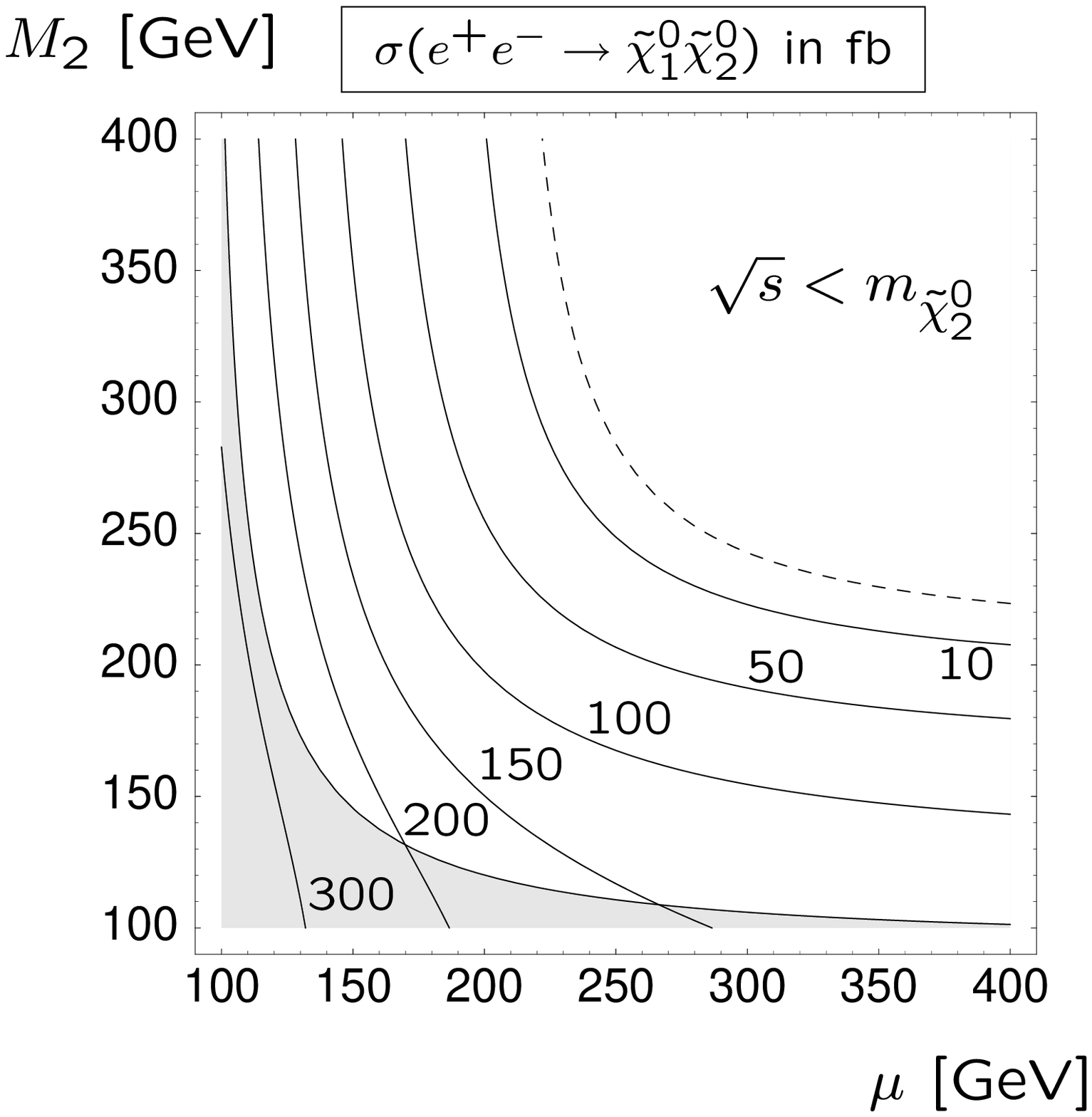}}
\put(210,-20){\includegraphics{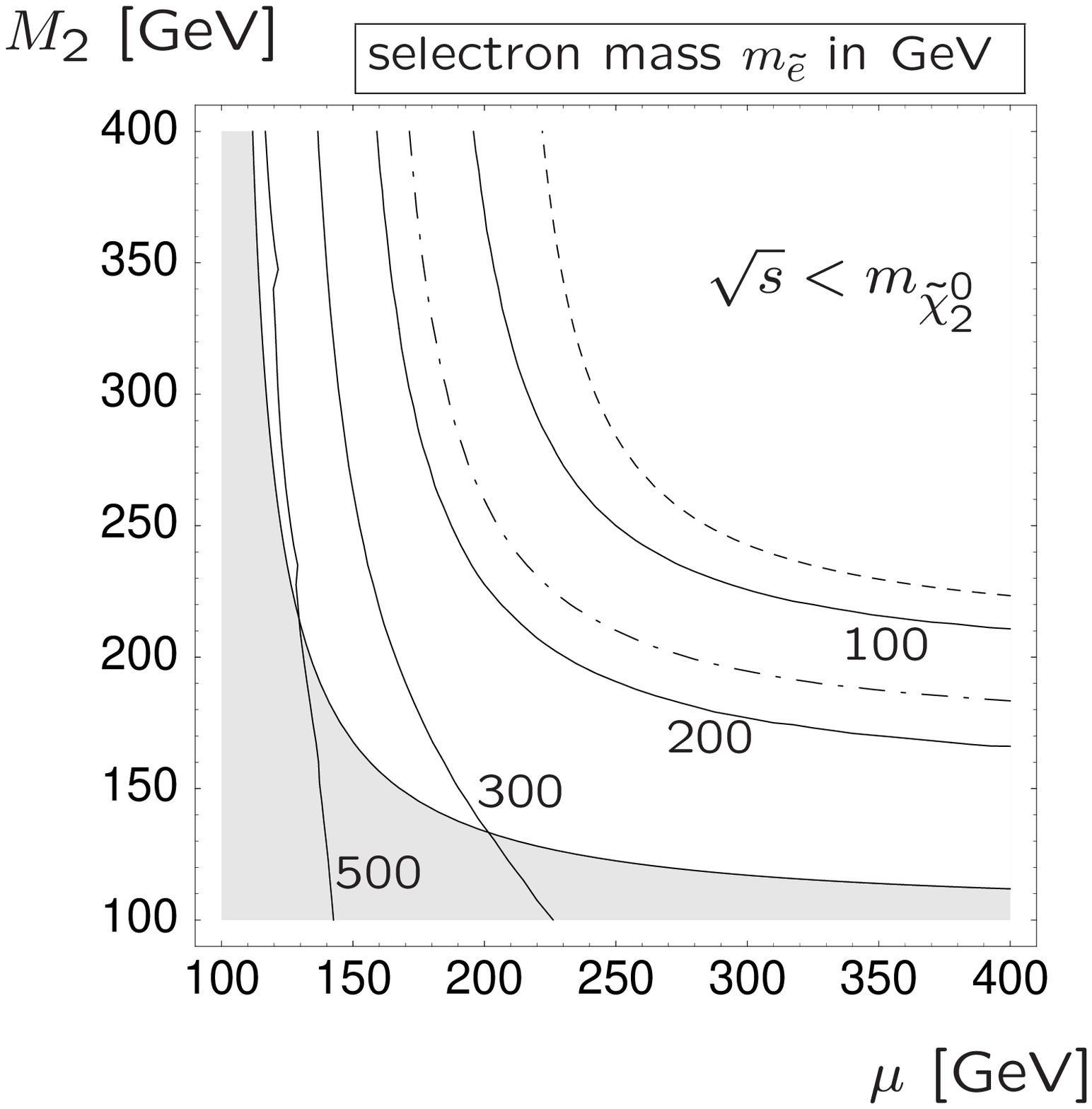}}
\put(40,10){\small\bf (a)}
\put(260,10){\small\bf (b)}
\end{picture}
\caption{\small 
     {\bf (a)} Contour lines in the $\mu$--$M_2$ plane of
     the neutralino production cross section 
      $\sigma(e^+e^-\to\tilde\chi_1^0\tilde\chi_2^0)$
      with $\tan\beta=10$, and $m_{\tilde e_R}=m_{\tilde e_L}=
      m_{\tilde e}=200$~GeV, at $\sqrt s = 208$~GeV.  At each point, $M_1$
      is chosen such that $m_{\tilde\chi_1^0}=0$.  
      {\bf (b)} Contour lines in the
      $\mu$--$M_2$ plane of the lower bounds on the selectron mass
       $m_{\tilde e_R}=m_{\tilde e_L}=m_ {\tilde e}$, such that
      $ \sigma(e^+e^-\to\tilde\chi_1^0\tilde\chi_2^0) =70$~fb 
      for $m_{\tilde\chi_1^0}=0$ with $\tan\beta=10$.  
      In {\bf (a}), {\bf (b)}, the dashed lines
      indicate the kinematical limit $m_{\tilde\chi_2^0}= \sqrt s=208$~GeV, 
      in the gray shaded areas the chargino mass is 
      $m_{\tilde\chi_1^\pm}<94$~GeV.  
      Along the dot-dashed contour in {\bf (b)} the relation
      $m_{\tilde e}=m_{\tilde\chi_2^0} $ holds.
}
\label{fig:OPALbounds}
\end{figure}

\textbf{Neutralino production at LEP:} 
If we assume $m_{\tilde\chi_1^0} =0$, the associated production 
$e^+e^-\to \tilde\chi_1^0 \tilde\chi_2^0$ would be accessible
at LEP up to the kinematical limit of $\sqrt s= m_{\tilde\chi_2^0} = 208$~GeV.
In order to compare with the results of the LEP searches we make use
of the model-independent upper bounds on the topological neutralino
production cross section obtained by OPAL with $\sqrt s=
208$~GeV~\cite{Abbiendi:2003sc},
\begin{equation}
\sigma(e^+e^-\to  \tilde\chi_1^0 \tilde\chi_2^0 )\times
{\rm BR}(\tilde\chi_2^0  \to Z\tilde\chi_1^0)\times {\rm BR}(Z\to q\bar q).
\end{equation}
Taking into account ${\rm BR}(Z\to q\bar q)\approx70\%$, one can
roughly read off from the 
OPAL\footnote{We analyze this bound assuming 
           conservatively that ${\rm BR}(\tilde\chi_2^0 \to Z\tilde\chi_1^0) = 1$.
           Note that OPAL has only considered the
           hadronic $Z$ decay channel, $Z\to q \bar q$. 
           If other leptonic neutralino decays  open, for example 
           2-body (or 3-body) decays via sleptons, see the dot-dashed line
           in Fig.~\ref{fig:OPALbounds}(b), this would lead
           to a reduction of the hadronic signal OPAL searched 
           for, and thus would allow for higher neutralino
           production cross sections. In that sense our bounds
           on these production cross sections are conservative.} 
plots~\cite{Abbiendi:2003sc},
\begin{equation}
\sigma(e^+e^-\to \tilde\chi_1^0 \tilde\chi_2^0 ) 
\times {\rm BR}(\tilde\chi_2^0 \to Z\tilde\chi_1^0 ) <70\,\mathrm{fb}.  
\label{bound}
\end{equation}
This is already a very tight bound, since typical 
neutralino production cross sections can be of the
order of $100$~fb, see Fig.~\ref{fig:OPALbounds}(a).
 For bino-like neutralinos, the main contribution to the cross section is due to
$\tilde e_R$ exchange.
Thus, the bound on the neutralino production
cross section can be translated into lower bounds on the selectron
mass $m_{\tilde e}$, for $m_{\tilde\chi_1^0}=0$.  
In Fig.~\ref{fig:OPALbounds}(b), we show lower bounds of the selectron mass,
such that along the contours the bound
 $\sigma(e^+e^-\to \tilde\chi_1^0 \tilde\chi_2^0)=70$~fb is fulfilled. 

\medskip

\textbf{Radiative neutralino production:}
Another search channel at LEP is radiative neutralino production,
$e^+e^-\to\tilde\chi_1^0\tilde\chi_1^0\gamma$. However,
due to the large SM background from radiative neutrino production
$e^+e^-\to\nu\bar\nu\gamma$, we find that the significance is always
$S<0.1$ for $\mathcal L=100$~pb$^{-1}$
and $\sqrt s= 208$~GeV~\cite{Dreiner:2006sb,Dreiner:2007vm}. Cuts on
the photon energy or angle do not help, due to similar distributions
of signal and background. 
At the ILC however, radiative neutralino production will be measurable,
due to a higher luminosity and the option of polarized 
beams~\cite{Dreiner:2006sb,Dreiner:2007vm,MoortgatPick:2005cw}.

%Since the OPAL analysis assumes heavy sleptons,
%the shown bounds are (conservative) upper bounds.
%The bounds could be tightened allowing for smaller sleptons masses.

%%%%%%%%%%%%%%%%%%%%%%%%%%%%%%%%%%%%%%%%%%%%%%%%%%%%%%%%%%%%%%%%%%%%%%%%%%%%%%
%%%%%%%%%%%%%%%%%%%%%%%%%%%%%%%%%%%%%%%%%%%%%%%%%%%%%%%%%%%%%%%%%%%%%%%%%%%

\section{Bounds from precision observables and rare decays}
\medskip

\begin{figure}[t]
\begin{picture}(200,150)
\put(-30,0){\includegraphics{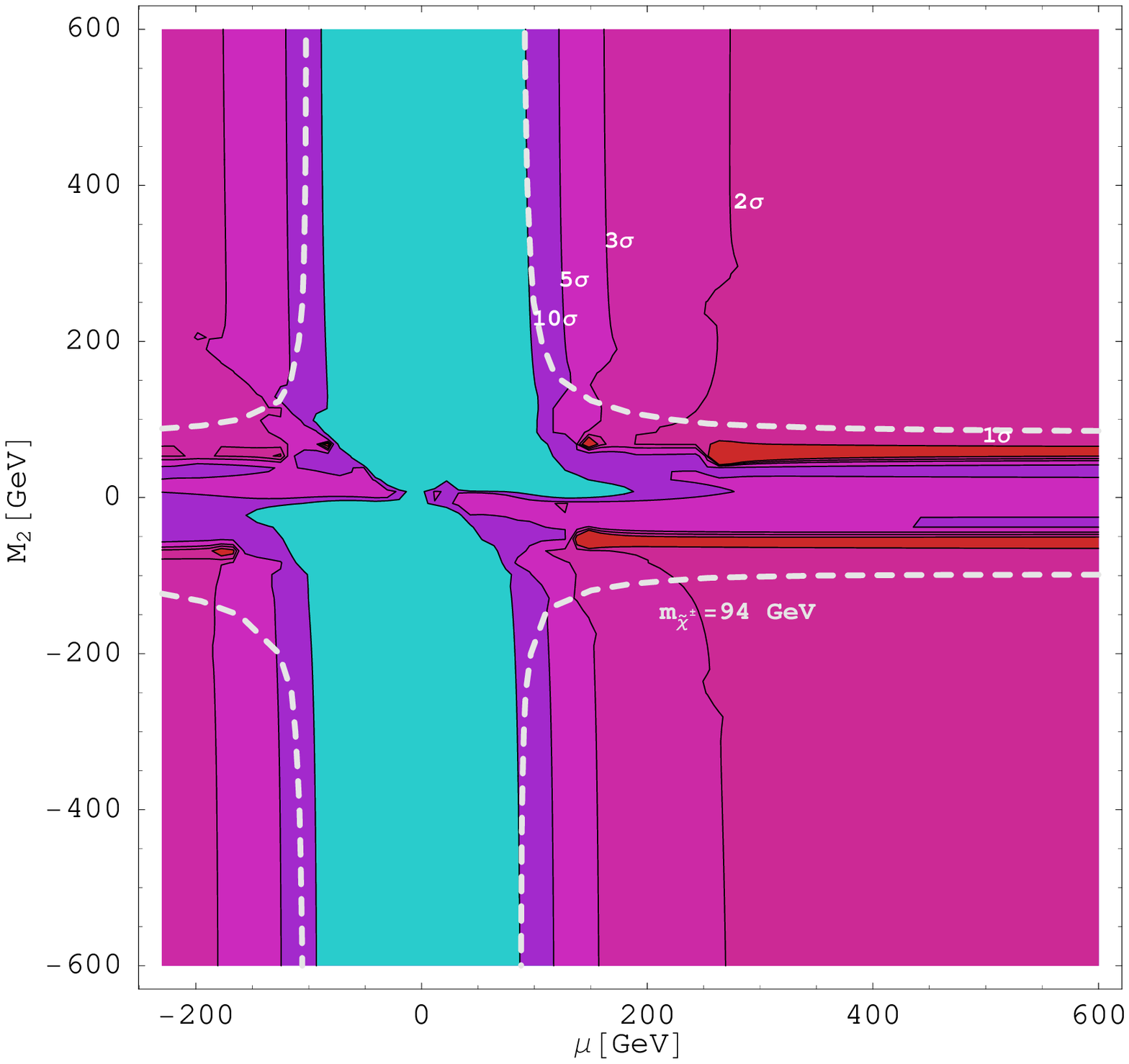}}
\put(210,0){\includegraphics{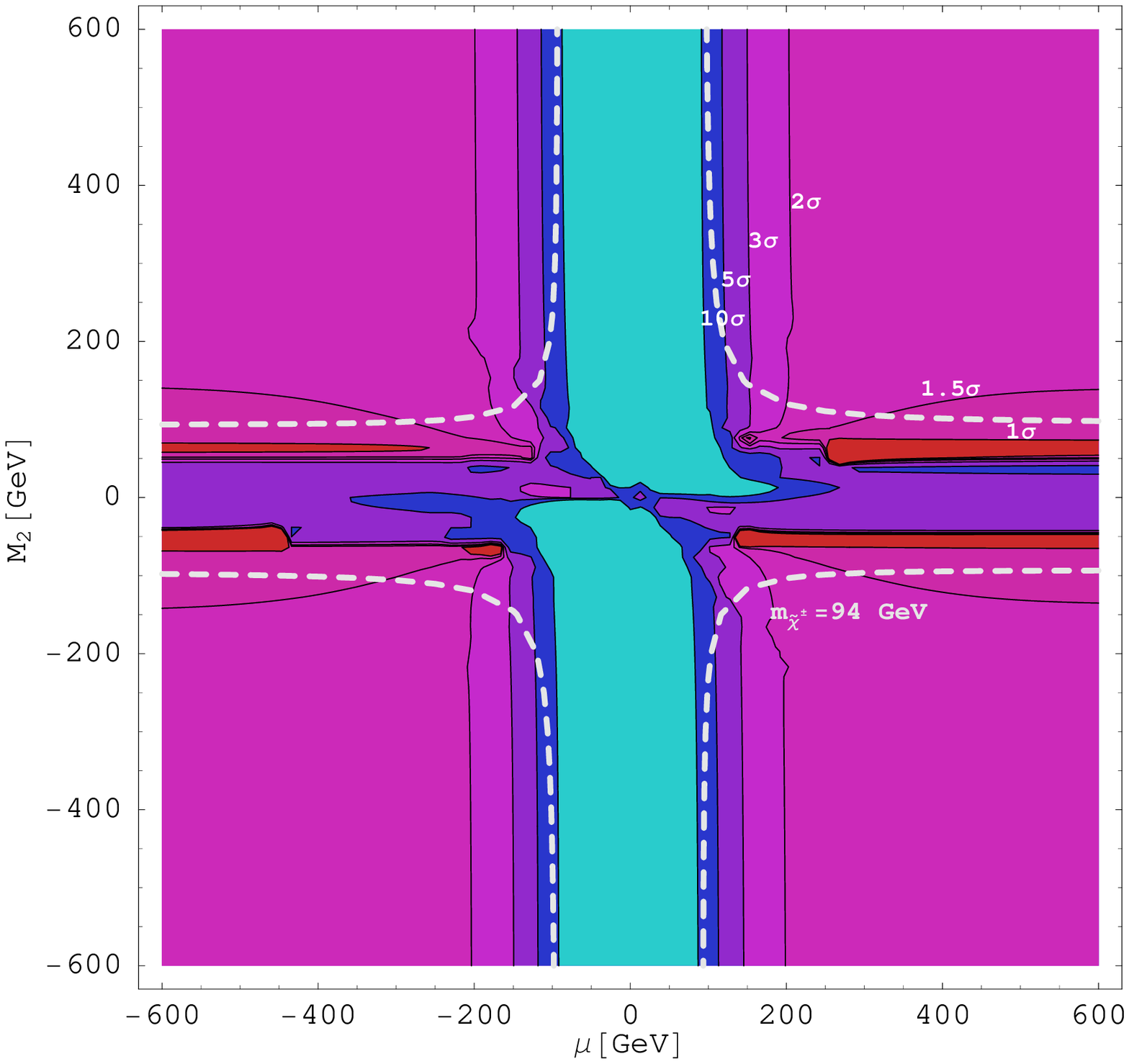}}
\put(20,-5){\small\bf (a)}
\put(255,-5){\small\bf (b)}
\end{picture}
\caption{\small 
         Contour lines in the $\mu$--$M_2$ plane for the difference 
         $\delta \Gamma =  
        (\Gamma_{\rm inv}-\Gamma_{\rm inv}^{\rm exp})/
        \Delta\Gamma$
         of theory prediction and
         experimental value of the invisible $Z$ width
         in units of the
         experimental error $\Delta\Gamma = 1.5~{\rm MeV}$,
         for  $m_{\tilde\chi_1^0} = 0$~GeV,
         $\tan\beta=10$, and {\bf (a)}
         $A_\tau=A_t=A_b=m_{\tilde g}=M_A=2M_{\tilde f}=500$~GeV,
          {\bf (b)}
          $A_\tau=A_t=A_b=m_{\tilde g}=M_A=M_{\tilde f}=600$~GeV.
          Along the dashed line 
          $m_{\tilde\chi_1^\pm}=94$~GeV.
}
\label{fig:invZwidth}
\end{figure}

In the following we study the impact of a light or massless neutralino
on electroweak precision physics. As an example, we focus on the
invisible $Z$ width, $\Gamma_{\rm inv}$, which
is potentially very sensitive to a light or massless neutralino,
due to the contribution $Z\to\tilde\chi_1^0\tilde\chi_1^0$, which
involves the higgsino contribution of the neutralino.
However, a light neutralino is mainly bino-like for $|\mu|\gsim125$~GeV,
see Fig.~\ref{fig:neutmixandmass}. 
In Fig.~\ref{fig:invZwidth}, we show the difference
$\delta \Gamma =  
        (\Gamma_{\rm inv}-\Gamma_{\rm inv}^{\rm exp})/
        \Delta\Gamma$
%$\delta \Gamma =  \Gamma_{\rm inv}-\Gamma_{\rm inv}^{\rm exp}$
from the measured invisible width
$\Gamma_{\rm inv}^{\rm exp} = 499.0\pm 
1.5$~MeV~\cite{Amsler:2008zz,lepewwg},
in units of the experimental
error $\Delta\Gamma=1.5$~MeV,
to the theoretical prediction $\Gamma_{\rm inv}$.
The calculations of $\Gamma_{\rm inv}$
include the full ${\mathcal O}(\alpha)$ SM and
MSSM contributions, supplemented with leading 
higher-order terms~\cite{Heinemeyer:2007bw}.
The deviation from the measured width
$\Gamma_{\rm inv}^{\rm exp}$ is larger than $5\sigma$
only for $|\mu|\lsim125$~GeV, where an increasing 
higgsino admixture leads to a non-negligible neutralino coupling to the $Z$. 
However those parts of the $\mu$-$M_2$~planes are mostly already 
excluded by direct chargino searches at LEP.
Note also that already the SM contribution to $\Gamma_{\rm inv}$
is more than $1\sigma$ larger than the experimental value 
$\Gamma_{\rm inv}^{\rm exp}$~\cite{lepewwg,Heinemeyer:2007bw}.

%%%%%%%%%%%%%%%%%%%%%%%%%%%%%%%%%%%%%%%%%%%%%%%%%%%%%%%%%%%%%%%%%%%%%%%%%%%%%%
%%%%%%%%%%%%%%%%%%%%%%%%%%%%%%%%%%%%%%%%%%%%%%%%%%%%%%%%%%%%%%%%%%%%%%%%%%%

%\medskip
\newpage
A massless or light neutralino has low impact on the $W$ boson mass, 
the effective leptonic weak mixing  angle $\sin^2\theta_{\rm eff}$, 
the electric dipole moments of the electron, neutron and mercury, 
and the anomalous magnetic moment of the muon $(g-2)_\mu$. We thus 
refer the reader to the original paper~\cite{Dreiner:2009ic}.
Rare meson decays into a light bino-like neutralino have also been 
analyzed~\cite{Dreiner:2009er}, but  no constraints on the neutralino
 mass could be set.

\section{Bounds from astrophysics and cosmology}
\medskip

\textbf{Supernova cooling:} 
Light neutralinos of masses of order $100$~MeV
could be thermally produced inside a Supernova.
If their mean free path is of the order of the
Supernova core size or lager, the neutralinos escape
freely and lead to an additional cooling of the 
Supernova~\cite{Dreiner:2003wh,Grifols:1988fw,Ellis:1988aa}.
To be in agreement with observations of the Kamiokande and IMB Collaborations
from SN 1987A, see Ref.~\cite{Dreiner:2003wh},
the cooling must not shorten the neutrino signal.
The energy that is emitted by the neutralinos
is much smaller than that emitted by the neutrinos if 
$m_{\tilde\chi_1^0} \gsim 200$~MeV~\cite{Dreiner:2003wh},
with $m_{\tilde e}=500$~GeV.
For heavy sleptons,
$m_{\tilde e}\gsim1200$~GeV, however, 
no bound on  the neutralino mass 
can be set~\cite{Dreiner:2003wh}.

\medskip

\textbf{Hot dark matter:} 
We consider the case of a nearly massless neutralino,
$ m_{\tilde\chi_1^0} \lsim \mathcal{O}(1\, \mathrm{eV})$. Since the very light bino
contributes to the hot dark matter of the universe, we assume here
implicitly that the cold dark matter originates from another
source. The bino relic energy density, $\rho_{\bino}$,
divided by the critical energy density of the universe, $\rho_c$, is
given by \cite{Kolb:1990eu}
\begin{eqnarray}
\label{ul:eq:relicdensity}
\Omega_{\bino} &\equiv& \frac{\rho_{\bino}}{\rho_c}
\;=\;  \frac{43}{11}\,\zeta(3)\,\frac{8\pi G_N}
{3 H_0^2}\,\frac{g_{\mathrm{eff}}(\bino)}{g_{\ast S}(T)}\,
            T_\gamma^3 \,m_{\bino}\,.
\end{eqnarray}
In order for the bino hot dark matter not to disturb the large structure
formation, we assume its contribution to be less than the upper bound
on the energy density of the neutrinos, as determined by the WMAP 
data~\cite{Dunkley:2008ie}
\begin{equation}
\label{eq:bound}
\Omega_{\bino} h^2 \le  [\Omega_{\nu} h^2]_{\mathrm{max}} = 0.0076\,.
\end{equation}
{}From Eqs.~(\ref{ul:eq:relicdensity}) and (\ref{eq:bound}), we 
find the conservative upper bound
\begin{eqnarray}
m_{\bino}  \le 0.7 \enspace\mathrm{eV}\,.
\end{eqnarray}
Thus a very light bino with mass below about $1$~eV is consistent
with structure formation. This line of argument was originally used by
Gershtein and Zel'dovich~\cite{Gershtein:1966gg} and Cowsik and
McClelland~\cite{Cowsik:1972gh} to derive a neutrino upper mass bound,
by requiring $\Omega_\nu\leq1$. We have here obtained an upper
mass bound for a hot dark matter bino.

\medskip

\textbf{Cold dark matter:} 
The impact of a light neutralino on its thermal relic density
has widely been studied~\cite{cosmology}.
If the neutralino accounts for the dark matter, 
its mass has to be $m_{\tilde\chi_1^0} > 3\dots20$~GeV.
Although seeming theoretically unmotivated,
those bounds could in principle be evaded by allowing 
a small amount of R-parity violation~\cite{Choudhury:1999tn},
and/or additional dark matter candidates.

\medskip

Note that many authors have revisited the case of a light WIMP in the sub
$10$~GeV mass range, to explain recent results from the DAMA/LIBRA, 
CDMS and/or CoGeNT experiments~\cite{Bottino:2009km}. 
In the MSSM, to ensure their effective annihilation,
 such particles must exchange a light $\mathcal{O}({\rm GeV})$ 
pseudoscalar Higgs at large values of $\tan\beta$. Since this is however ruled 
out by recent TEVATRON results on SUSY Higgs searches, the authors of 
Ref.~\cite{Vasquez:2010ru} recently concluded that a light 
MSSM neutralino of $m_{\tilde\chi_1^0} < 15$~GeV should be excluded.
In Ref.~\cite{Feldman:2010ke} it was pointed out that also improved measurements 
on $B_s\to \mu\mu$ exclude neutralinos with such light masses to accommodate
the CoGeNT preferred region.

\ack

I would like to thank Herbi Dreiner, Sven Heinemeyer, Ulrich Langenfeld, and
Georg Weiglein for the collaborations underlying this work. This work was
partially supported by SFB TR-33 The Dark Universe, and
 by MICINN project FPA.2006-05294.

\section*{References}
%lala

%\end{footnotesize}


\begin{thebibliography}{99}

%--------SUSY--------------------
\bibitem{Haber:1984rc}
  H.~E.~Haber and G.~L.~Kane,
%  ``The Search For Supersymmetry: Probing Physics Beyond The Standard Model,''
  Phys.\ Rept.\  {\bf 117}, 75 (1985).
  %%CITATION = PRPLC,117,75;%%


%-----------Proton Hexality

\bibitem{Dreiner:2005rd}
  H.~Dreiner, C.~Luhn and M.~Thormeier,
  {\em Phys. Rev.}  D {\bf 73} (2006) 075007
  [arXiv:hep-ph/0512163].
  %%CITATION = PHRVA,D73,075007;%%

\bibitem{Dreiner:2007vp}
  H.~Dreiner, et al.
  {\em Nucl. Phys.} {\bf B 795} (2008) 172
  [arXiv:0708.0989 [hep-ph]].
  %%CITATION = NUPHA,B795,172;%%


%%-- previous own work on light neutralino


%\cite{Choudhury:1999tn}
\bibitem{Choudhury:1999tn}
  D.~Choudhury, H.~K.~Dreiner, P.~Richardson and S.~Sarkar,
  %``A supersymmetric solution to the KARMEN time anomaly,''
  Phys.\ Rev.\  D {\bf 61}, 095009 (2000); \\
%%X  [arXiv:hep-ph/9911365].
  %%CITATION = PHRVA,D61,095009;%%
%\cite{Dedes:2001zia}
%\bibitem{Dedes:2001zia}
  A.~Dedes, H.~K.~Dreiner and P.~Richardson,
  %``Attempts at explaining the NuTeV observation of dimuon events,''
  Phys.\ Rev.\  D {\bf 65}, 015001 (2001)
  [arXiv:hep-ph/0106199].
  %%CITATION = PHRVA,D65,015001;%%


%\cite{Dreiner:2003wh}
\bibitem{Dreiner:2003wh}
  H.~K.~Dreiner, C.~Hanhart, U.~Langenfeld and D.~R.~Phillips,
  %``Supernovae and light neutralinos: SN1987A bounds on supersymmetry
  %revisited,''
  Phys.\ Rev.\  D {\bf 68}, 055004 (2003)
  [arXiv:hep-ph/0304289].
  %%CITATION = PHRVA,D68,055004;%%


%\cite{Dreiner:2006sb}
\bibitem{Dreiner:2006sb}
  H.~K.~Dreiner, O.~Kittel and U.~Langenfeld,
  %``Discovery potential of radiative neutralino production at the ILC,''
  Phys.\ Rev.\  D {\bf 74} (2006) 115010
  [arXiv:hep-ph/0610020].
  %%CITATION = PHRVA,D74,115010;%%


%\cite{Dreiner:2007vm}
\bibitem{Dreiner:2007vm}
  H.~K.~Dreiner, O.~Kittel and U.~Langenfeld,
  %``The role of beam polarization for radiative neutralino production at the
  %ILC,''
  {\em Eur. Phys. J.}  {\bf C 54} (2008) 277
  [arXiv:hep-ph/0703009].
  %%CITATION = HEP-PH/0703009;%%



\bibitem{Dreiner:2009ic}
  H.~K.~Dreiner, S.~Heinemeyer, O.~Kittel, U.~Langenfeld, A.~M.~Weber and G.~Weiglein,
  %``Mass Bounds on a Very Light Neutralino,''
  Eur.\ Phys.\ J.\  C {\bf 62}, 547 (2009)
  [arXiv:0901.3485 [hep-ph]];
  %%CITATION = EPHJA,C62,547;%%
%
%\cite{Dreiner:2007fw}
%\bibitem{Dreiner:2007fw}
%  H.~K.~Dreiner, S.~Heinemeyer, O.~Kittel, U.~Langenfeld, A.~M.~Weber and G.~Weiglein,
  %``How light can the lightest neutralino be?,''
%{\it In the Proceedings of 2007 International Linear Collider Workshop (LCWS07 and ILC07), Hamburg, Germany, 30 May - 3 Jun 2007, pp SUS06}
  arXiv:0707.1425 [hep-ph].
  %%CITATION = ECONF,C0705302,SUS06;%%

%\cite{Dreiner:2009yk}
\bibitem{Dreiner:2009yk}
  H.~K.~Dreiner,
  %``Comments on a Massless Neutralino,''
  AIP Conf.\ Proc.\  {\bf 1200}, 73 (2010)
  [arXiv:0910.1509 [hep-ph]].
  %%CITATION = APCPC,1200,73;%%




%%X%\cite{Hooper:2002nq}
%%X\bibitem{Hooper:2002nq}
\bibitem{cosmology}
%%X%\cite{Bednyakov:1996ax}
%%X\bibitem{Bednyakov:1996ax}
%%X  V.~A.~Bednyakov, H.~V.~Klapdor-Kleingrothaus and S.~G.~Kovalenko,
%%X  %``Superlight neutralino as a dark matter particle candidate,''
%%X  Phys.\ Rev.\  D {\bf 55}, 503 (1997)
%%X  [arXiv:hep-ph/9608241];
%%X  %%CITATION = PHRVA,D55,503;%%
  D.~Hooper and T.~Plehn,
  %``Supersymmetric dark matter - how light can the LSP be?,''
  Phys.\ Lett.\  B {\bf 562}, 18 (2003)
  [arXiv:hep-ph/0212226];\\
  %%CITATION = PHLTA,B562,18;%%
%%X
%
%%\cite{Belanger:2002nr}
%\bibitem{Belanger:2002nr}
  G.~Belanger, F.~Boudjema, A.~Pukhov and S.~Rosier-Lees,
%  %``A lower limit on the neutralino mass in the MSSM with non-universal gaugino
%  %masses. ((T)) ((U)),''
  arXiv:hep-ph/0212227;\\
%  %%CITATION = HEP-PH/0212227;%%
%
%%X%%\cite{Bottino:2002ry}
%%X\bibitem{Bottino}
  A.~Bottino, N.~Fornengo and S.~Scopel,
  %``Light relic neutralinos,''
  Phys.\ Rev.\  D {\bf 67}, 063519 (2003)
  [arXiv:hep-ph/0212379];\\
  %%CITATION = PHRVA,D67,063519;%%
%%\cite{Bottino:2003iu}
%\bibitem{Bottino:2003iu}
  A.~Bottino, F.~Donato, N.~Fornengo and S.~Scopel,
  %``Lower bound on the neutralino mass from new data on CMB and  implications
  %for relic neutralinos,''
  Phys.\ Rev.\  D {\bf 68}, 043506 (2003)
  [arXiv:hep-ph/0304080]; \\
  %%CITATION = PHRVA,D68,043506;%%
%% %\cite{Bottino:2004qi}
%% \bibitem{Bottino:2004qi}
%%  A.~Bottino, F.~Donato, N.~Fornengo and S.~Scopel,
%%  %``Indirect signals from light neutralinos in supersymmetric models  without
%%  %gaugino mass unification,''
  Phys.\ Rev.\  D {\bf 70}, 015005 (2004)
  [arXiv:hep-ph/0401186]; \\
  %%CITATION = PHRVA,D70,015005;%%
%%X
%%X
%%X%\cite{Belanger:2003wb}
%%X\bibitem{Belanger:2003wb}
  G.~Belanger, F.~Boudjema, A.~Cottrant, A.~Pukhov and S.~Rosier-Lees,
  %``Lower limit on the neutralino mass in the general MSSM. ((V),''
  JHEP {\bf 0403}, 012 (2004)
  [arXiv:hep-ph/0310037];\\
  %%CITATION = JHEPA,0403,012;%%
%\cite{Lee:2007ai}
%\bibitem{Lee:2007ai}
  J.~S.~Lee and S.~Scopel,
  %``Lightest Higgs boson and relic neutralino in the MSSM with CP violation,''
  Phys.\ Rev.\  D {\bf 75}, 075001 (2007)
  [arXiv:hep-ph/0701221];\\
  %%CITATION = PHRVA,D75,075001;%%
%\cite{Hooper:2008au}
%\bibitem{Hooper:2008au}
  D.~Hooper, T.~Plehn and A.~Vallinotto,
  %``Neutralino Dark Matter and Trilepton Searches in the MSSM,''
  Phys.\ Rev.\  D {\bf 77}, 095014 (2008)
  [arXiv:0801.2539 [hep-ph]].
  %%CITATION = PHRVA,D77,095014;%%


%-new refs light neutralino------------------

%\cite{Bottino:2009km}
\bibitem{Bottino:2009km}
  A.~Bottino, F.~Donato, N.~Fornengo, S.~Scopel,
  %``Relic neutralinos and the two dark matter candidate events of the CDMS II
  %experiment,''
  Phys.\ Rev.\  D {\bf 81} (2010) 107302
  [arXiv:0912.4025 [hep-ph]]; \\
  %%CITATION = PHRVA,D81,107302;%%
%
%\cite{Kuflik:2010ah}
%\bibitem{Kuflik:2010ah}
  E.~Kuflik, A.~Pierce and K.~M.~Zurek,
  %``Light Neutralinos with Large Scattering Cross Sections in the Minimal
  %Supersymmetric Standard Model,''
  Phys.\ Rev.\  D {\bf 81}, 111701 (2010)
  [arXiv:1003.0682 [hep-ph]];\\
  %%CITATION = PHRVA,D81,111701;%%
%
%\cite{Fitzpatrick:2010em}
%\bibitem{Fitzpatrick:2010em}
  A.~L.~Fitzpatrick, D.~Hooper and K.~M.~Zurek,
  %``Implications of CoGeNT and DAMA for Light WIMP Dark Matter,''
  Phys.\ Rev.\  D {\bf 81}, 115005 (2010)
  [arXiv:1003.0014 [hep-ph]];\\
  %%CITATION = PHRVA,D81,115005;%%
%
%\cite{Hisano:2009xv}
%\bibitem{Hisano:2009xv}
  J.~Hisano, K.~Nakayama and M.~Yamanaka,
  %``Implications of CDMS II result on Higgs sector in the MSSM,''
  Phys.\ Lett.\  B {\bf 684}, 246 (2010)
  [arXiv:0912.4701 [hep-ph]];\\
  %%CITATION = PHLTA,B684,246;%%
%
%\cite{Asano:2009kj}
%\bibitem{Asano:2009kj}
  M.~Asano, S.~Matsumoto, M.~Senami and H.~Sugiyama,
  %``CDMS II result and Light Higgs Boson Scenario of the MSSM,''
  JHEP {\bf 1007}, 013 (2010)
  [arXiv:0912.5361 [hep-ph]].
  %%CITATION = JHEPA,1007,013;%%




%\cite{Vasquez:2010ru}
\bibitem{Vasquez:2010ru}
  D.~A.~Vasquez, G.~Belanger, C.~Boehm, A.~Pukhov and J.~Silk,
  %``Can neutralinos in the MSSM and NMSSM scenarios still be light?,''
  arXiv:1009.4380 [hep-ph].
  %%CITATION = ARXIV:1009.4380;%%

%\cite{Feldman:2010ke}
\bibitem{Feldman:2010ke}
  D.~Feldman, Z.~Liu and P.~Nath,
  %``Low Mass Neutralino Dark Matter in the MSSM with Constraints from $B_s\to
  %\mu^+\mu^-$ and Higgs Search Limits,''
  Phys.\ Rev.\  D {\bf 81} (2010) 117701
  [arXiv:1003.0437 [hep-ph]].
  %%CITATION = PHRVA,D81,117701;%%

%-----------PDG-----------------lala
\bibitem{Amsler:2008zz}
  C.~Amsler et al.\  [Particle Data Group],
  {\em Phys. Lett.}  {\bf B 667} (2008) 1.
  %%CITATION = PHLTA,B667,1;%%


%\cite{Gogoladze:2002xp}
\bibitem{Gogoladze:2002xp}
  I.~Gogoladze, J.~D.~Lykken, C.~Macesanu and S.~Nandi,
  %``Implications of a massless neutralino for neutrino physics,''
  Phys.\ Rev.\  D {\bf 68}, 073004 (2003).
%%X  [arXiv:hep-ph/0211391].
  %%CITATION = PHRVA,D68,073004;%%



%\cite{Abbiendi:2003sc}
\bibitem{Abbiendi:2003sc}
  G.~Abbiendi {\it et al.}  [OPAL Collaboration],
  %``Search for chargino and neutralino production at s**(1/2) = 192-GeV to
  %209-GeV at LEP,''
  Eur.\ Phys.\ J.\  C {\bf 35} (2004) 1
  [arXiv:hep-ex/0401026].
  %%CITATION = EPHJA,C35,1;%%

%\cite{MoortgatPick:2005cw}
\bibitem{MoortgatPick:2005cw}
  G.~A.~Moortgat-Pick {\it et al.},
  %``The role of polarized positrons and electrons in revealing fundamental
  %interactions at the linear collider,''
  arXiv:hep-ph/0507011.
  %%CITATION = HEP-PH/0507011;%%

\bibitem{lepewwg} The ALEPH, DELPHI, L3, OPAL, SLD Collaborations,
                   the LEP Electroweak Working Group, \\
                   the SLD Electroweak and Heavy Flavour Groups,
                   hep-ex/0509008;
%                    \\
                    %%CITATION = HEP-EX 0509008;%%
%                   {}[The ALEPH, DELPHI, L3 and OPAL Collaborations, the LEP
%                   Electroweak Working Group], 
                   hep-ex/0612034;
                   %%CITATION = HEP-EX 0612034;%%
%
%\bibitem{LEPEWWG} 
                   LEP Electroweak Working Group,
                   see: {\tt http://lepewwg.web.cern.ch/LEPEWWG/Welcome.html}.

%\cite{Heinemeyer:2007bw}
\bibitem{Heinemeyer:2007bw}
  S.~Heinemeyer, W.~Hollik, A.~M.~Weber and G.~Weiglein,
  %``$Z$ Pole Observables in the MSSM,''
  JHEP {\bf 0804} (2008) 039
  [arXiv:0710.2972 [hep-ph]].
  %%CITATION = JHEPA,0804,039;%%

%--------MESON deays-----------------------------------------
%\cite{Dreiner:2009er}
\bibitem{Dreiner:2009er}
  H.~K.~Dreiner, S.~Grab, D.~Koschade, M.~Kramer, B.~O'Leary and U.~Langenfeld,
  %``Rare meson decays into very light neutralinos,''
  Phys.\ Rev.\  D {\bf 80}, 035018 (2009)
  [arXiv:0905.2051 [hep-ph]].
  %%CITATION = PHRVA,D80,035018;%%



%------------supernova cooling
\bibitem{Grifols:1988fw}
  J.~Grifols, E.~Masso and S.~Peris,
  {\em Phys. Lett.}  {\bf B 220} (1989) 591.
  %%CITATION = PHLTA,B220,591;%%

%
\bibitem{Ellis:1988aa}
  J.~Ellis, K.~Olive, S.~Sarkar and D.~Sciama,
  {\em Phys. Lett.}  {\bf B 215}, 404 (1988).
  %%CITATION = PHLTA,B215,404;%%

%------------Hot dark matter

\bibitem{Kolb:1990eu}
  E.~Kolb and M.~Turner
	``{\it The Early Universe}''
	(Westview Press 1990).



\bibitem{Dunkley:2008ie}
  J.~Dunkley et al.\  [WMAP Collaboration],
  arXiv:0803.0586 [astro-ph].
  %%CITATION = ARXIV:0803.0586;%%

%
\bibitem{Gershtein:1966gg}
  S.~Gershtein and Y.~Zeldovich,
  {\em JETP Lett.\ } {\bf 4} (1966) 120
  [{\em Pisma Zh.\ Eksp.\ Teor.\ Fiz.\ } {\bf 4} (1966) 174].
  %%CITATION = ZFPRA,4,174;%%


%\cite{Cowsik:1972gh}
\bibitem{Cowsik:1972gh}
  R.~Cowsik and J.~McClelland,
  %``An Upper Limit on the Neutrino Rest Mass,''
  Phys.\ Rev.\ Lett.\  {\bf 29} (1972) 669.
  %%CITATION = PRLTA,29,669;%%


\end{thebibliography}
\end{document}